\documentclass[prl,twocolumn,amsmath,amssymb,showkeys,showpacs]{revtex4}

\usepackage{graphicx}
\usepackage{dcolumn}
\usepackage{bm}

\newcommand {\omegarf}{\omega_{r\!f}}
\newcommand {\frf}{f_{r\!f}}
\newcommand {\Irf}{I_{r\!f}}
\newcommand {\jrf}{\jmath_{r\!f}}

\newcommand {\jdc}{\jmath_{d\!c}}

\begin{document}

\title{Parallel pumping of magnetic vortex gyrations in spin-torque nano-oscillators}

\date{\today}

\author{P. Bortolotti} \altaffiliation{Corresponding author. Electronic address: paolo.bortolotti@thalesgroup.com}
\affiliation{Unit\'e Mixte de Physique CNRS/Thales and Universit\'e Paris Sud 11, 1 av. Fresnel, 91767 Palaiseau, France}
\author{C. Serpico} \affiliation{Dipartimento di Ingegneria Elettrica, Universit\'a di Napoli Federico II, Via Claudio 21, Napoli, Italy}
\author{E. Grimaldi} \affiliation{Unit\'e Mixte de Physique CNRS/Thales and Universit\'e Paris Sud 11, 1 av. Fresnel, 91767 Palaiseau, France}
\author{A. Dussaux} \altaffiliation{Present address: ETH Zurich, 101 R\"amistrasse, 8092 Zurich, Switzerland} \affiliation{Unit\'e Mixte de Physique CNRS/Thales and Universit\'e Paris Sud 11, 1 av. Fresnel, 91767 Palaiseau, France}
\author{J. Grollier} \affiliation{Unit\'e Mixte de Physique CNRS/Thales and Universit\'e Paris Sud 11, 1 av. Fresnel, 91767 Palaiseau, France}
\author{V. Cros} \affiliation{Unit\'e Mixte de Physique CNRS/Thales and Universit\'e Paris Sud 11, 1 av. Fresnel, 91767 Palaiseau, France}
\author{K. Yakushiji} \affiliation{Institute of Advanced Industrial Science and Technology (AIST), Spintronics Research Center, Tsukuba, Japan}
\author{A. Fukushima} \affiliation{Institute of Advanced Industrial Science and Technology (AIST), Spintronics Research Center, Tsukuba, Japan}
\author{H. Kubota} \affiliation{Institute of Advanced Industrial Science and Technology (AIST), Spintronics Research Center, Tsukuba, Japan}
\author{R. Matsumoto} \affiliation{Institute of Advanced Industrial Science and Technology (AIST), Spintronics Research Center, Tsukuba, Japan}
\author{S. Yuasa} \affiliation{Institute of Advanced Industrial Science and Technology (AIST), Spintronics Research Center, Tsukuba, Japan}

\pacs{75.47.-m, 75.40.Gb, 85.75.-d, 85.30.Mn}

\keywords{parametric excitation, parallel pumping, spin transfer oscillator, magnetic vortex dynamics}

\begin{abstract}
We experimentally demonstrate that large magnetic vortex oscillations can be parametrically excited in a magnetic tunnel junction by the injection of radio-frequency ($r\!f$) currents at twice the natural frequency of the gyrotropic vortex core motion. The mechanism of excitation is based on the \emph{parallel pumping} of vortex motion by the $r\!f$ orthoradial field generated by the injected current. Theoretical analysis shows that experimental results can be interpreted as the manifestation of \emph{parametric amplification} when $r\!f$ current is small, and of \emph{parametric instability} when $r\!f$ current is above a certain threshold. By taking into account the energy nonlinearities, we succeed to describe the amplitude saturation of vortex oscillations as well as the coexistence of stable regimes.
\end{abstract}	

\maketitle

Parametric excitations of magnetization oscillations
have been extensively studied in the area of ferromagnetic
resonance \cite{Lvov-1994} in connection with the processes of
spin-wave instability \cite{Suhl-1956} and parallel pumping \cite{Schloemann-1962}.
These phenomena are driven by the modulation of the oscillation frequency
of elementary excitations (spin-waves) associated to the spatially uniform
ground state  obtained by a strong bias field.
In the case of spin-wave instability, the spatially uniform mode  is driven
to large precession angle by radio-frequency (rf) fields
applied transversal to the bias field. The modulation of spin-wave
spectrum is due to the nonlinear coupling of the uniform precession with spin-waves.
On the other hand, in the case of parallel pumping,
the rf field is applied along the direction of the bias field,
and thus directly modulates the frequency of spin-wave
modes \cite{Schloemann-1962}.

In this letter, we investigate both experimentally and theoretically the implementation of
the principle of {\em parallel pumping} in spin-torque devices having a vortex in the free magnetic layer. It is an unconventional case of parallel pumping because the field used to modulate the frequency of magnetization oscillations is the orthoradial magnetic field generated by the injected current (usually referred as Oersted field). The frequency of this modulation is chosen to be twice the frequency of the lowest frequency mode present in a magnetic thin disk with vortex ground state, which is the translational motion of the vortex core \cite{Ivanov-2002}.

The present investigation is relevant to the development and control of  spintronics
rf-oscillators in large excitation regimes~\cite{Slavin-2008}. These oscillators promise to have an important role in near future microwave communication technologies~\cite{Ralph-2008,Villard-2010}.
A striking result in this areas has been the development of spin-torque nano-oscillators (STNO),
tunable over a wide frequency range by the injected currents both for uniform~\cite{Kiselev-2003,Rippard-2004,Sankey-2006} and non-uniform~\cite{Pribiag-2007,Mistral-2008,Puffal-2007} magnetic configurations. The performances of these devices in terms of power and linewidth have been constantly improved over the last few years~\cite{Krivorotov-2008,Georges-2009,Georges-2008, Quinsat-2010,Keller-2010,Pogoryelov-2011,Gusakova-2011}. In particular, it has been recently demonstrated~\cite{Dussaux-2010,Khvalkovskiy-2009,Dussaux-2012} that Magnetic Tunnel Junctions (MTJ) pillars with vortex ground state~\cite{Guslienko-2002,Guslienko-2008,Gaididei-2010} allow to obtain signal with large output power ($\simeq 1\mu$W) and very good coherence ($\simeq 1$ MHz).
It is crucial to gain understanding of the dynamics of these oscillators in regimes where the vortex is driven very far from the equilibrium position as it is the case of the present study.

Parametric resonance by means of parallel pumping in a spin-valve nanopillar has been previously
studied by \emph{Urazhdin et al.}~\cite{Urazhdin-2010}
in the more traditional setting where the magnetic free layer is uniformly magnetized and the pumping
rf field, generated by an antenna, is applied along the spatially uniform bias field. In addition, the somehow related phenomena of phase-locking of spin-torque auto-oscillations by an external source at double frequency, 
has been investigated both in the case of uniformly magnetized free layer~\cite{Urazhdin-2010}, and in the case of vortex state free layer~\cite{Martin-2011}.

Here, we investigate the parametric excitation in a vortex-MTJ subject
to an external $r\!f$ current with a frequency close to $2f_0$, where $f_0$ is
the gyrotropic frequency of the vortex core with a dc-current below the threshold for self-oscillations (sub-critical case) used to reduce dissipative forces on the vortex. 
The device is subject to a strong out-of-plane static field, used to tilt the magnetization in the polarizing
layer, which leads to an out-of-plane component of the vortex curling in the free layer \cite{Ivanov-2002}.
The parallel pumping is realized by modulating the gyrotropic frequency of the vortex
through the rf variations of the Oersted field. The resulting gyrotropic motion exhibits two specific features of parametric resonance of weakly dissipative oscillators. The first effect, usually referred as \emph{parametric amplification} (see e.g.~\cite{Rugar-1991}), consists in the amplification of signals which contains the frequencies close to the natural oscillation frequency of the vortex. In our case, it manifests itself in the amplification of the thermally  activated vortex motion around its equilibrium position. The second effect is the \emph{parametric instability} that takes place when the power of the $r\!f$ excitation exceeds a certain threshold at which both the amplitude and the coherence of the oscillations strongly increase. Eventually saturation is reached due to nonlinearities.

\begin{figure}[]
\begin{center}
\includegraphics[width=8cm]{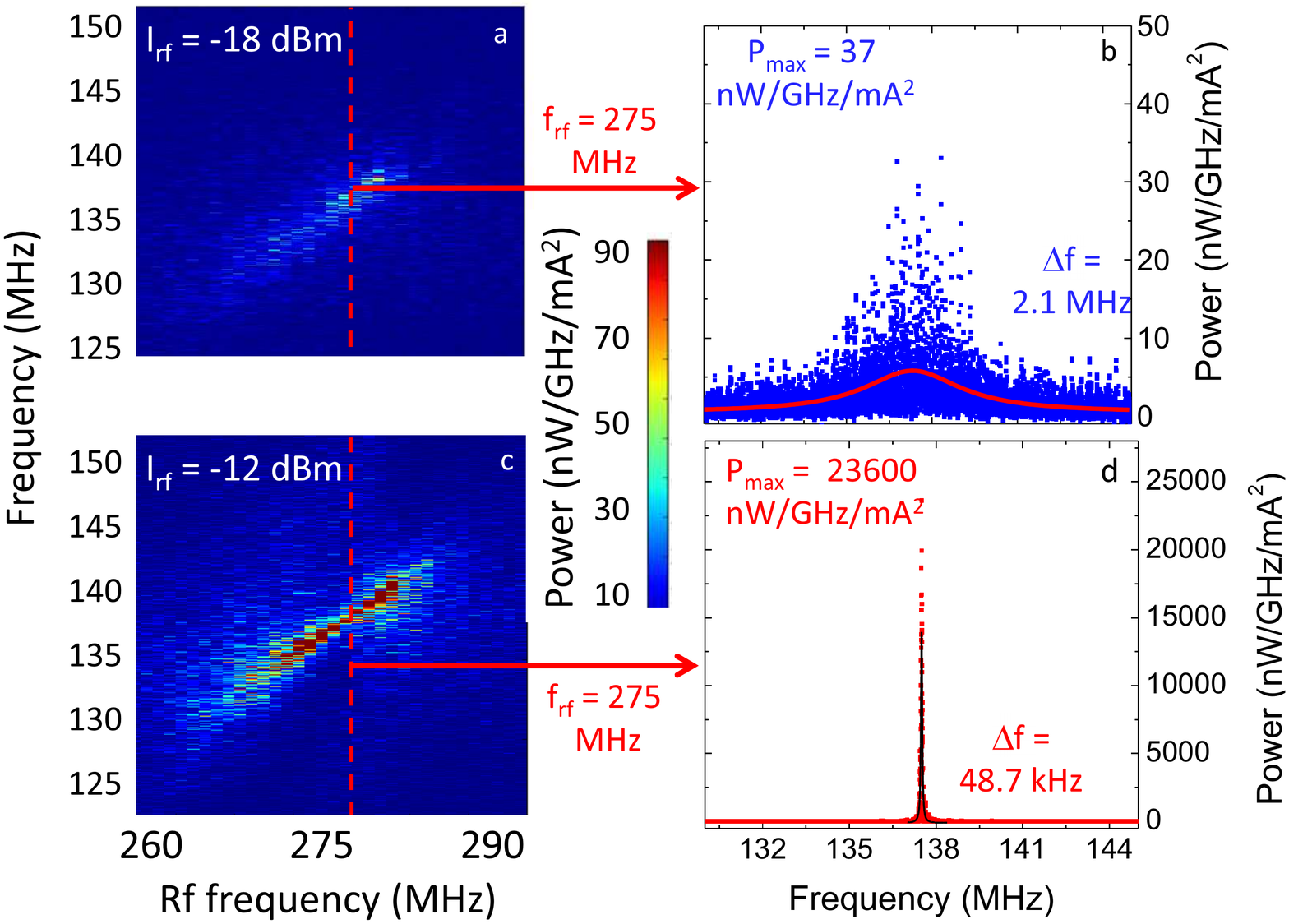}
\end{center}
\caption{\label{fig:Spectra} a-c) Colored maps of emitted power for the frequency of the gyrotropic core motion \emph{versus} the external $r\!f$ frequency measured (a) for a small ($-18$ dBm) and (c) a large $r\!f$ current ($-12$ dBm). $I_{dc}=3$ mA and $H_{perp}=4.48$ kG. b-d) Power spectral density \emph{versus} frequency measured with an external $r\!f$ current at $f_{r\!f}=275$ MHz.}
\end{figure}

We perform our measurements on several circular MTJs of 500 nm diameter. The complete structure (with thickness in nm) is: PtMn(15) / CoFe(2.5) / Ru(0.85) / CoFeB(3) / MgO(1.075) / NiFe(5) / Ta(7) / Ru(6) / Cr(5) / Au(200). The $5$ nm-thick NiFe free layer presents a vortex magnetization at remanence.
The MTJ resistance at the saturated state is $R \simeq 45 \, \Omega$ with average magnetoresistance $\Delta R \simeq 8.5 \, \Omega$ at room temperature. For the whole presented measurements, a dc current $I_{dc} = +3$ mA and an out-of-plane field $H_{ext,z}=4.48$ kG, are applied.
In our convention, a positive current means electrons flowing from the free layer to the SAF stack. We emphasize that the injected dc current $I_{dc}$ is always below the threshold $I_{th}$ necessary to excite large amplitude vortex-core sustained oscillations. Consequently, when no $I_{r\!f}$ current is applied, the small $I_{dc}$ induces only small oscillations of the vortex core and a very weak signal is measured at $f_0 \simeq 138$ MHz with maximum power about $3$ nW/GHz/mA$^2$ and a large linewidth (about $12$ MHz). These figures are typical of thermal-induced vortex core oscillations~\cite{Dussaux-2010,Bortolotti-2012,Petit-2007}. The main features of the microwave signal are strongly modified when an additional current $I_{r\!f}$ is injected at a frequency close to $2f_0$. In Fig.~\ref{fig:Spectra}a, we first present the color-scale map of the power spectral density (PSD) as a function of the frequency of the external $r\!f$ current at a small power equal to $-18$ dBm (corresponding to $I_{r\!f} = 0.8$ mA). A signal emerges from the background level when $264 < f_{r\!f} < 278$ MHz, which corresponds to a frequency around twice the natural frequency $f_0$ of the system. As shown in Fig.\ref{fig:Spectra}b, the spectrum measured for $f_{r\!f} =275$ MHz can be fitted by a Lorentzian function and a maximum power of $37$ nW/GHz/mA$^2$, about one order of magnitude larger than the thermal one, is obtained. The corresponding linewidth is also reduced of one order of magnitude to $2.1$ MHz. To our knowledge, these measurements are the first observation of parametric \emph{amplification} for the vortex dynamics in the regime of sub-critical $I_{dc}$ currents and are consistent with the ones presented by \emph{Urazhdin et al.}~\cite{Urazhdin-2010} for the case of uniform magnetization under a parametric $r\!f$ field.

The situation changes drastically when $I_{r\!f}>-18$ dBm. We enter in the parametric \emph{instability} regime and a tremendous improvement for both the coherence and the emitted power is obtained. A striking example of such behavior is displayed in Fig.\ref{fig:Spectra}c for large $r\!f$ power equal to $-12$ dBm (corresponding to $I_{r\!f} = 1.59$ mA). We find a large window of $f_{r\!f}$, i.e., between $265$ and $285$ MHz, in which the emitted power associated to the vortex dynamics comes out from the background level and the spectral linewidth reduces significantly. In this range, the vortex frequency increases linearly with $f_{r\!f}$ and is strictly equal to $f_{r\!f}/2$. To demonstrate the drastic improvement of $r\!f$ features in this specific regime, we plot in Fig.~\ref{fig:Spectra}d, the peak detected for $f_{r\!f}=275$ MHz. The measured linewidth $\Delta f$ of $49$ kHz is indeed limited by the resolution bandwidth (RBW) used for such large range of measured frequency. Additional measurements with optimized RBW, allow us to extract a bottom limit value of the linewidth equal to $9$ kHz. The maximum power reaches $23.6 \mu$W/GHz/mA$^2$ that is three orders of magnitude larger than the case of low $r\!f$ power.

In order to describe the phenomena observed in experiments, we consider the Thiele equation for the description of translational motion of the vortex core~\cite{Guslienko-2008,Dussaux-2010} in the free layer of the MTJ device. This equation can be written as:
\begin{equation}\label{eq:DynSTVNO_X}
    \bm G \times \frac{d \bm X}{dt} + D \frac{d \bm X}{dt} =
    -\frac{\partial W}{\partial \bm X}  + \bm F_{st,z} + \bm F_{ef\!f} \,
\end{equation}
where $\bm X=(X_1,X_2)$ is the vortex core position. The first term at the left-hand-side (LHS) of Eq.~\eqref{eq:DynSTVNO_X} is the gyrotropic term  with $\bm G= -\bm e_z 2\pi  L {M_s }/{\gamma} $ ~\cite{Guslienko-2002} ($\gamma$ the gyromagnetic ratio, $L$ is layer thickness, and $M_s$ the saturation magnetization). The unit vector $\bm e_z$ is perpendicular to the plane of the layer and directed from the polarizer to the free layer. The second term at the LHS of Eq.~\eqref{eq:DynSTVNO_X} is the damping $D=\alpha \eta |\bm G|$ where $\alpha$ is the Landau-Lifshitz damping, and $\eta\approx 1$ is a geometric factor. The first term at the right-hand-side (RHS) of Eq.~\eqref{eq:DynSTVNO_X} is the force associated to the gradient of the vortex energy $W(\bm X)$. This terms is given by the sum of the vortex magnetostatic energy $W_{ms}$, and the vortex Zeeman energy $W_{oe}$ associated to the Oersted field which is proportional to the  injected current density $\jmath=I/(\pi R^2)= \jrf \cos(2\pi \frf t) + \jmath_{dc}$ ($R$ is the pillar radius). The expressions of these two energy terms, which introduce the main nonlinearity in the system along with the coupling with the RF excitations, are $W_{oe}=\lambda_{oe}\jmath X^2 + \lambda^\prime_{oe}\jmath X^4/R^2$ and $W_{ms}=-(\kappa_{ms}R^2/2) \log(1-[X/2R]^2)$, where $\lambda_{oe}=0.85 \mu_0 M_s RL$, $\lambda_{oe}^\prime=-0.5\lambda_{oe}$ and $\kappa_{ms}=({10}/{9})\mu_0 M_s^2{L^2}/{R}$~\cite{Dussaux-2012,Gaididei-2010,Guslienko-2002}. The term $ \bm F_{st,z}=C_z p_z j  \, \bm e_z \times \bm X$ is the spin transfer force responsible to compensate the damping forces. It depends on the perpendicular component of the polarizer magnetization $p_z=\bm p \cdot \bm e_z$ and on $C_z=|\bm G|\gamma \sigma/2$, a constant measuring the spin-torque efficiency, where $\sigma= {\hbar P}/({2|e| L M_s})$, $P$ is the spin polarization, $\hbar$ the Planck constant and $e$ the electron charge. The term $\bm F_{ef\!f}$ at the RHS of Eq.~\eqref{eq:DynSTVNO_X} takes into account all additional forces acting on the vortex. This terms includes spin-torque forces due to the in-plane component of the polarizer magnetization~\cite{Dussaux-2012}, static in-plane forces acting on the vortex core (for example, the remaining stray field from the SAF polarizer) and extra forces due, for example, to defects~\cite{Kim-2010,Tobik-2012}. Such force $\bm F_{ef\!f}$ is assumed to be approximately independent of $\bm X$. The main effect of $\bm F_{ef\!f}$ under dc excitation is to produce a break of the rotational symmetry of the system which leads to a static shift of the core position from the MTJ center to the position $\bm X_0$. The effect of this force is equivalent to an effective in-plane field $H_{ef\!f,xy}$. Even if the precise determination of $H_{ef\!f,xy}$ is not easy to obtain, from resistance \emph{versus} in-plane field measurements, we estimate it to be few tens of Gauss, thus enough to break the system symmetry even at low injected $r\!f$ currents. Finally, the strong out-of-plane external field is taken into account
by an appropriate rescaling of the parameters entering in all terms of Eq.~\eqref{eq:DynSTVNO_X} \cite{Ivanov-2002,Dussaux-2012}.

In order to study vortex oscillation around the displaced position, it is convenient to use the normalized core displacement $\delta \bm x= (\bm X -\bm X_0)/R$. After appropriate algebraic manipulations~\cite{Serpico-2013}, 
and by neglecting inessential small terms, Eq.~\eqref{eq:DynSTVNO_X} can be rewritten as follows:
\begin{equation}\label{eq:ddeltax_general}
\begin{split}
   \frac{d}{dt}\, \delta \bm x  =  A_0(t)\cdot \delta \bm x+ \bm N\left(t,\delta \bm x \right)
   + \bm v(t)  \, ,
\end{split}
\end{equation}
where
\begin{equation}\label{eq:A0(t)}
   A_0(t)= \begin{pmatrix}
                     -d \widetilde{\omega}_1(t) +c_z p_z \jmath_{dc}   & -\widetilde{\omega}_2(t) \\
                     \widetilde{\omega}_1(t)  &  -d \widetilde{\omega}_2(t) + c_z p_z \jmath_{dc} \\
         \end{pmatrix} \, ,
\end{equation}
$d=D/|\bm G|$ is the normalized damping, $c_z= C_z/|\bm G|$ is the normalized spin-torque efficiency and
the two angular frequencies
\begin{equation}\label{eq:parametric_freq}
  \widetilde{\omega}_{1,2}(t)= \omega_{1,2} \left[ 1+ q_{1,2} \jrf \cos(2\pi \frf \, t)\right] \,
\end{equation}
are related to the quadratic approximation $W(\delta \bm x) \approx  [\widetilde{\omega}_{1}(t)\delta x_1^2 + \widetilde{\omega}_{2}(t)\delta x_2^2]/2$ of the vortex energy around $\bm X_0$. In Eq.~\eqref{eq:parametric_freq},
$\omega_1= \Omega_{dc} + \Omega^\prime_{dc}x_0^2$, $\omega_2= \Omega_{dc} + 3\Omega^\prime_{dc}x_0^2$, where $x_0=|\bm X_0|/R$, $\Omega_{dc}= (\kappa_{ms}+\lambda_{oe}\jdc)/ |\bm G|$
 and $\Omega_{dc}^\prime= (0.25 \kappa_{ms}+\lambda^\prime_{oe}\jdc)/ |\bm G|$. The frequency $f_0$ of vortex
 gyrations around $\bm X_0$ is given by the formula $\omega_0=2\pi f_0 = \sqrt{\omega_1 \omega_2}$. The parameters $q_{1}$ and $q_{2}$ are related to the $r\!f$ component of $W_{oe}$: 
 $q_1=(\lambda_{oe}+3\lambda_{oe}^\prime x_0^2)/(\omega_1 |\bm G|)$ and $q_2=(\lambda_{oe}+\lambda_{oe}^\prime x_0^2)/(\omega_2 |\bm G|)$ \cite{Serpico-2013}. The term $\bm N\left(t,\delta \bm x \right)=\bm N_0\left(\delta \bm x \right)\cos(\omegarf t)$ in Eq.~\eqref{eq:ddeltax_general} contains all nonlinear terms in $\delta {\bm x}$. Finally, the last term $\bm v(t)=\bm v_0 \cos(\omegarf t)$ is a forcing term independent of $\delta \bm x$,
which comes from the rf components of $\bm F_{ef\!f}$.

The fact that in general $\widetilde{\omega}_{1}(t) \neq \widetilde{\omega}_{2}(t)$ is a consequence of the break of the symmetry due to the force $\bm F_{ef\!f}$ and it is indeed this break that controls the coupling between vortex oscillation and parametric excitations. The situation is similar to the case of usual parallel pumping in rotationally symmetric systems. On that case, the coupling between the $2 f_0$ excitations and the magnetization oscillations is controlled by the ellipticity of the amplitude which is due to the dipolar fields generated by the spin-waves \cite{Schloemann-1962}. Instead, in our case parametric pumping of vortex dynamics takes place only if $q_1 \neq q_2$, which occurs when the vortex oscillates around a displaced position.

To treat analytically Eq.~\eqref{eq:ddeltax_general}, it is important to notice that conservative terms in the RHS are expected to be the dominant terms in vortex core oscillations. The other terms act as perturbations on the system and thus induce only slow changes of the vortex oscillation. As a consequence, in order to describe this slow variation when $\omegarf \approx 2\omega_0$, it is convenient to make the following (Van Der Pol type) change of variable $\delta {\bm x} \mapsto {\bm a}$:
\begin{equation}\label{eq:deltaxtoa_trasf}
  \delta \bm x(t) = \begin{pmatrix}
                     \cos(\omegarf \, t/2)   & -(1/u)\,\sin(\omegarf \, t/2) \\
                     u\sin(\omegarf \, t/2)  &  \cos(\omegarf \, t/2) \\
         \end{pmatrix}  \cdot \bm a(t)
\end{equation}
where $u=({\omega_1/\omega_2})^{1/2}$ and $\omegarf=2\pi \frf$. By substituting Eq.~\eqref{eq:deltaxtoa_trasf} into Eq.~\eqref{eq:ddeltax_general} and by taking the time-average of the resulting equation over one period of the $r\!f$ oscillation, one arrives to the following equation
\begin{equation}\label{eq:a_nlinav_eq}
 \frac{d\bm a}{dt}= \begin{pmatrix}
                     - d\, \bar{\omega}_0(a) +c_z p_z \jmath_{dc}  & -\delta\bar{\omega}_0(a) -\jrf \Gamma\\
                     \delta\bar{\omega}_0(a)-\jrf \Gamma &  - d\, \bar{\omega}_0(a) +c_z p_z \jmath_{dc} \\
         \end{pmatrix} \cdot \bm a
\end{equation}
where $\delta\bar{\omega}_0(a)= \bar{\omega}_0(a) - \omegarf/2$ is the nonlinear detuning parameter and
\begin{eqnarray}\label{eq:omegarbara}
\bar{\omega}_0(a) = \omega_0 + \nu_{oe}j_{dc} a^2 + \Omega_{ms}[1/(1-(a/2)^2)] \, ,
\end{eqnarray}
with $\Omega_{ms}= \kappa_{ms}/|\bm G|$ and $ \nu_{oe}=\lambda_{oe}/|\bm G|$. The parameter which control the coupling with the $r\!f$ excitations is $\jrf \Gamma=\omega_0(q_2-q_1)/4$ which is zero if there is no shift of the vortex, i.e. $q_1=q_2$.

\begin{figure}[t]
\begin{center}
\includegraphics[width=8cm]{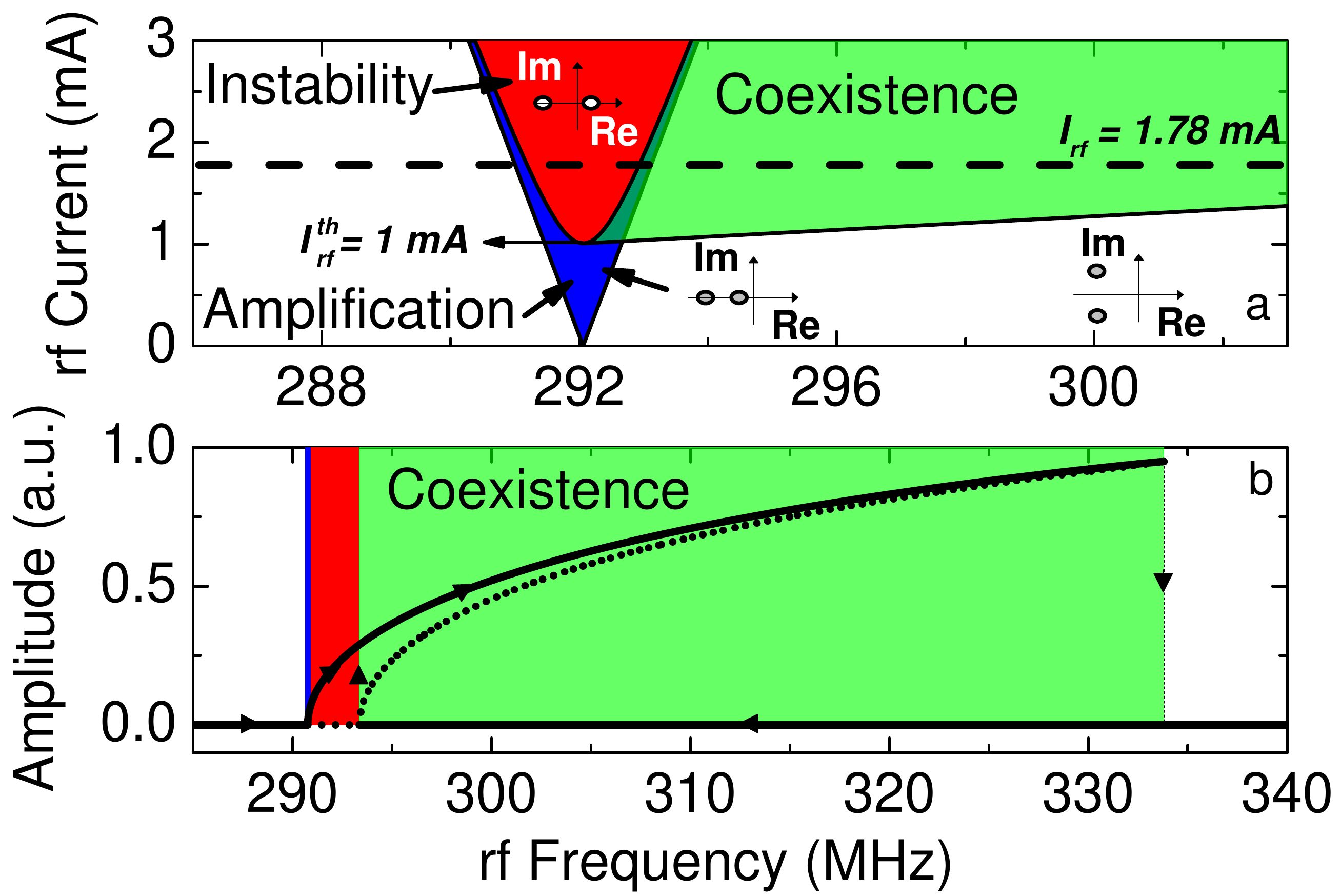}
\end{center}
\caption{\label{fig:FrequencyResponse} a) Phase diagram in the $(\frf,\jrf)$-plane with
  indication of region of parametric instability, coexistence of stable regimes and parametric amplification.  b) Analytically computed amplitude response versus $\frf$, according to Eq.~\eqref{eq:nlin_regimes} with $\Irf=1.78$ mA
  which corresponds to $-11$ dBm rf power. The values of the parameters used are: $P=0.225$, $I_{dc}=2.5$ mA, $\alpha=0.01$ (LL damping) and $x_0=0.3$.}
\end{figure}

Note that, in the condition under study, i.e. $I_{dc}<I_{th}$, there is no sustained oscillation and it is the $r\!f$ time-varying Oersted field contribution which is responsible to destabilize the vortex core from its equilibrium position. The stability of small magnetization oscillations around the static equilibrium position can be studied by considering the linearized version of Eq.~\eqref{eq:a_nlinav_eq} around $\bm a=0$. The eigenvalues of such linearized equation are given by
\begin{equation}\label{eq:eingenlinaveqa}
  \lambda_{1,2}=  - d\, {\omega}_0 +c_z p_z\jmath_{dc} \pm \sqrt{\jrf^2 \Gamma^2 -({\omega}_0 - \omegarf/2)^2} \, .
\end{equation}
The stability is controlled by the sign of the real part of $\lambda_{1,2}$ and the instability boundary in the $(\frf,\jrf)$-plane correspond to the condition of vanishing of one eigenvalue:
\begin{equation}\label{eq:hyperbola_inst}
   \jrf^2 \Gamma^2 -({\omega}_0 - \omegarf/2)^2=  (d\, {\omega}_0 -c_z p_z\jmath_{dc})^2 \, .
\end{equation}
This is the equation of the hyperbola centered at the point $(\omegarf=2\omega_0, \jrf=0)$ and with asymptotes along the lines $\jrf = \pm (\omegarf/2-{\omega}_0)/\Gamma$ (see Fig.~\ref{fig:FrequencyResponse}a). From Eq.~\eqref{eq:hyperbola_inst}, one derives that the threshold value of $\jrf$ such that instability occurs is
\begin{equation}\label{eq:threshold}
\jrf^{th}= (d\, {\omega}_0 -c_z p_z\jmath_{dc})/ \Gamma \, .
\end{equation}
For the particular parameter choice of Fig.~\ref{fig:FrequencyResponse}a, $\jrf^{th}=1$ mA. Below $\jrf^{th}$ we find the very same behavior observed in Fig.~\ref{fig:Spectra}a-b. In general, outside the hyperbola the real parts of $\lambda_{1,2}$ are negative and small vortex oscillation are stable. Nevertheless, if one consider linear response of the system to noise, one may observe two different qualitative behaviors. At relatively large detuning, the eingenvalues are complex conjugates and the frequency response of $\bm a$ has peaks at the frequencies $\frf/2-f_0$ and $f_0 -\frf/2$. By using Eq.~\eqref{eq:deltaxtoa_trasf}, i.e., going back to the stationary frame, one can infer that thermally driven vortex dynamics have power spectral density doubly peaked at $f_0$ and $\frf-f_0$. The behavior changes close to $2f_0$ (blue region in Fig.~\ref{fig:FrequencyResponse}a) when the eigenvalues become both real. The frequency response of $\bm a$ is centered at zero, and thus, by using again Eq.~\eqref{eq:deltaxtoa_trasf}, one can conclude that the power spectral density of thermally driven oscillations is centered at $\frf/2$. The amplitude of the linear response increases progressively as $(\frf, \jrf)$ approaches the instability boundary and the signal linewidth shrinks to zero. This behavior is the so-called \emph{parametric amplification}. In our experimental investigations, the evidences of parametric amplification are seen in the increase of thermally driven vortex oscillations which are precursors of parametric instability, see Fig.\ref{fig:Spectra}(a-b).

\begin{figure}[t]
\begin{center}
\includegraphics[width=8cm]{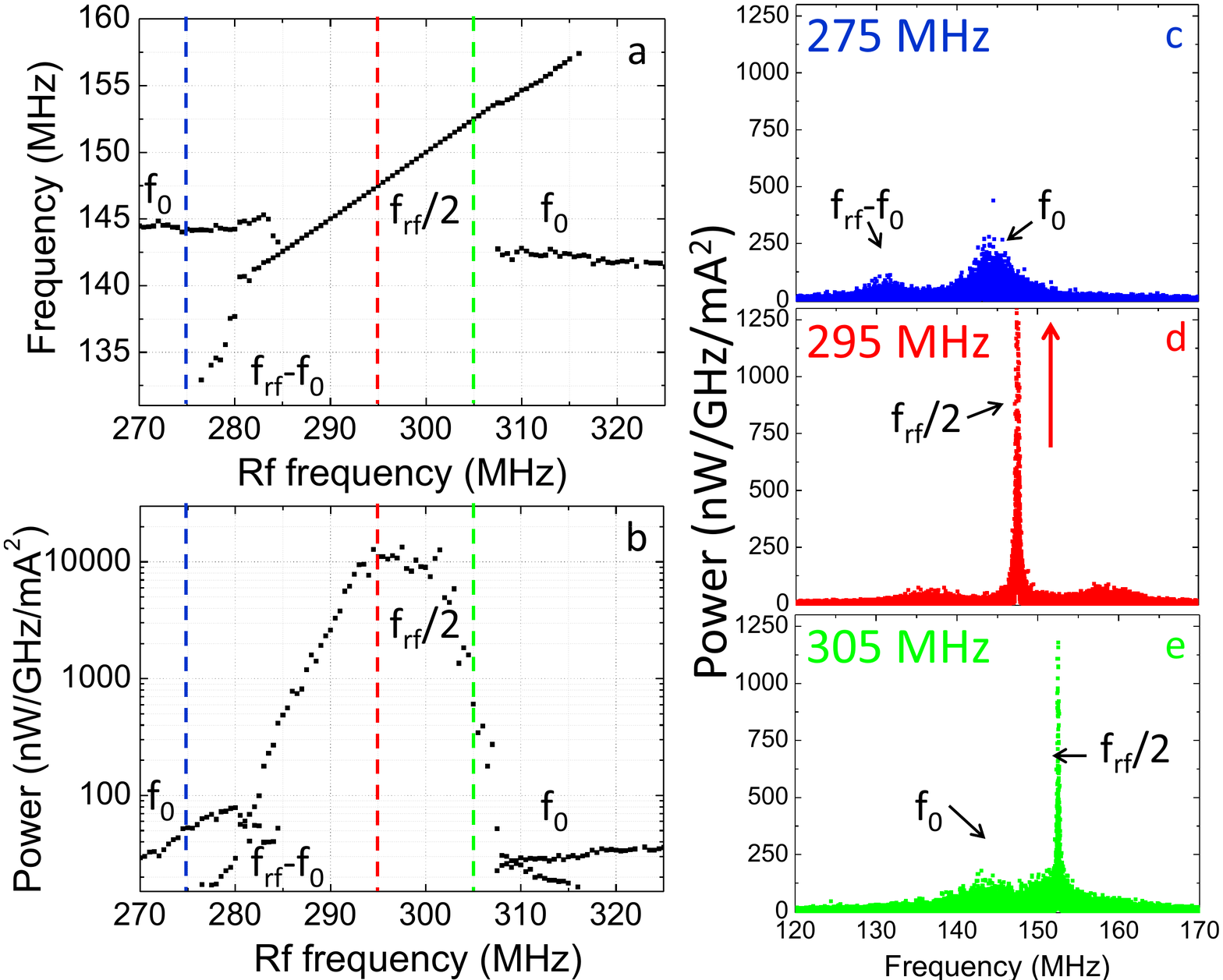}
\end{center}
\caption{\label{fig:NonLinearResponse} a) Vortex frequency $f$ \emph{versus} $r\!f$-frequency $f_{r\!f}$. b) Power density \emph{versus} $r\!f$-frequency $f_{r\!f}$. c-d-e) Power density \emph{versus} frequency $f$ for three values of $f_{r\!f}$ : $275$, $295$ and $305$ MHz. Note that these measurements have been made with an out-of-plane field $H_{perp}=4.48$ kG, a subcritical dc current $I_{dc}$ = 2.5 mA and a $r\!f$ current $I_{r\!f}=-11$ dBm.}
\end{figure}

Inside the instability region (red region in Fig.~\ref{fig:FrequencyResponse}), the equilibrium position of the vortex becomes unstable and the study of the system response requires the inclusion of nonlinear terms. The nonlinear regime after instability can be found by searching for nonzero equilibria of the general equation Eq.~\eqref{eq:a_nlinav_eq} which, by taking into account Eq.~\eqref{eq:deltaxtoa_trasf}, correspond to steady state of the vortex at a frequency $\frf/2$. This solution can be found by imposing that the determinant of the matrix at the RHS of Eq.~\eqref{eq:a_nlinav_eq} is zero, which leads to the equation:
\begin{equation}\label{eq:nlin_regimes}
  \jrf^2 \Gamma^2 -[\bar{\omega}_0(a) - \omegarf/2]^2=  [d\, \bar{\omega}_0(a) -c_z p_z\jmath_{dc}]^2 \, .
\end{equation}
The nonlinear response of the system can be then obtained by keeping $\jrf$ fixed and interpreting Eq.~\eqref{eq:nlin_regimes} as an implicit relation between $a$ and $\omega_{r\!f}$. The result of this analytical computation for the particular case of $I_{r\!f}=1.78$ mA (dashed black line in Fig.~\ref{fig:FrequencyResponse}a) is shown in Fig.~\ref{fig:FrequencyResponse}b. Solid lines indicate stable regimes, while the dashed lines indicate unstable regimes. In the white region only one state exists and it is stable ($\bm a=0$). In the red region, parametric excitation takes place and the amplitude of $\bm a$ start to increase while the state at $\bm a=0$ becomes unstable. In the region of coexistence (green region in Fig.~\ref{fig:FrequencyResponse}) the stable large amplitude regime coexists with the stable equilibrium position of the vortex.

To verify the validity of the analytical model, notably the prediction of the \emph{coexistence} (green region in Fig.~\ref{fig:FrequencyResponse}), we present in Fig.~\ref{fig:NonLinearResponse} a serie of measurements performed on an other junction from the same wafer, in which the vortex gyrotropic dynamics has a few MHz difference for the thermal resonant frequency. The measurements have been performed with $H_{perp}=4.48$ kG, $I_{dc}=2.5$ mA and $I_{r\!f}=-11$ dBm.

In Fig.~\ref{fig:NonLinearResponse}a-b, we plot the vortex frequency $f$ and the power of the emitted signal as a function of $f_{r\!f}$. At low frequencies ($f_{r\!f} < 275$ MHz), we detect only a small peak around $f_0$ corresponding to the thermally excited signal. At $f_{r\!f} = 275$ MHz, a second peak  with a very small power (note the logarithmic scale in Fig.~\ref{fig:NonLinearResponse}b) centered at $f_{r\!f}-f_0$ appears. This situation with two small peaks, whose power increases with $f_{r\!f}$ lasts until $285$ MHz. A typical spectrum in this regime is shown in Fig.~\ref{fig:NonLinearResponse}c. From $f_{r\!f} = 285$ until $305$ MHz, the spectra contains a main peak with large power and two side bands. The analysis of these thermally excited sideband signals in the parametric regime is out of the scope of this paper and for sake of clarity, we have not reported theirs values in Fig.~\ref{fig:NonLinearResponse}a-b. In this $r\!f$ frequency range, the vortex frequency is locked to the source signal, i.e., $f_{r\!f}/2$, and thus evolves linearly (see Fig.~\ref{fig:NonLinearResponse}a). As shown in Fig.~\ref{fig:NonLinearResponse}b, the peak power increases strongly to reach a maximum of $\approx 10 \mu$W/GHz/mA$^2$ at $f_{r\!f} = 295$ MHz. The linewidth obtained at this $r\!f$ frequency is only a few tens of kHz. Then, between $295$ and $300$ MHz, the power spectral density stagnates, meaning that the amplitude of the vortex oscillation saturates. Increasing further $f_{r\!f}$ above $300$ MHz, the power density start to decline. Moreover at $f_{r\!f} = 305$ MHz, we detect again a second peak centered at $f_{0}$ (see Fig.~\ref{fig:NonLinearResponse}e).

As expected by the analytical calculations (see Fig.~\ref{fig:FrequencyResponse}b), the two peaks, one at $f_{r\!f}/2$ and one at $f_{0}$, detected in the region between $305$ and $315$ MHz, correspond to the two states predicted in the region of coexistence. Notably, the main features of the power density evolution (see Fig.~\ref{fig:NonLinearResponse}b) is in very good agreement with the evolution expected by the theory (see Fig.~\ref{fig:FrequencyResponse}b): the oscillation amplitude grows monotonously from 100 to 10000 nW/GHz/mA$^2$ between $282$ and $295$ MHz. Conversely, we find that the opposite reduction of power occurs very sharply in few MHz (from $300$ to $305$ MHz). This behavior corresponds to the soft-hard regime discrimination already observed for the case of uniform magnetization~\cite{Urazhdin-2010}. Note that thermal effects, that are not considered in the model, result in a blur of the transition between the different regimes, and consequently allow some thermally induced transitions from one state to the other. Furthermore, we believe that another impact of thermal energy is to avoid the experimental observation of the hysteresis effect predicted in Fig.~\ref{fig:FrequencyResponse}b. Complementary experiments at low temperature and analysis of the parametrically driven vortex dynamics in time-domain might be helpful to address more precisely the impact of temperature on the non-linear vortex dynamics.

In summary, we have presented a comprehensive investigation of parametric excitation of vortex dynamics in a MTJ based spin transfer oscillator. Moreover, we propose an analytical model to predict the phase diagram of our vortex system in presence of an external $r\!f$ current at about twice the natural vortex frequency. We report the first observation in spin torque devices of the different parametric regimes of amplification and instability. Finally, we demonstrate both experimentally and theoretically, the coexistence of two parametric states of the vortex dynamics, evidencing the important role of non linearities. The parametric excitation in vortex based devices might be used for highly efficient $r\!f$ detection or low noise amplification using the specific potential of these devices in tuning their $r\!f$ features through both the $r\!f$ and $dc$ current.

The authors acknowledge N. Locatelli and N. Reyren for fruitful discussion, Y. Nagamine, H. Maehara and K. Tsunekawa of CANON ANELVA for preparing the MTJ films and the financial support from ANR agency (SPINNOVA  ANR-11-NANO-0016) and EU FP7 grant (MOSAIC No. ICT-FP7- n�317950). E.G. acknowledges DGA-CNES for financial support. C.S. acknowledges financial
support from  MIUR-PRIN 2010-11 Project2010ECA8P3 "DyNanoMag".


\clearpage
\onecolumngrid


\begin{thebibliography}{1}
\bibitem{Serpico-2013} Detailed derivations will be published elsewhere.
\end{thebibliography}
\end{document}